
\paperstyle{EQMAC}


\def\hc{\hbox{h.c.}}

\def\exp{\hbox{exp}}

\PACS 11.15Ha, 11.30.Rd, 11.30.Qc

\mithnumber 11 93

\paperdate April 18th 1993

\title Electroweak Interactions on a Planck Lattice:

\centerline{\bf the Propagators of Gauge Bosons}

\vfootnote{$\dagger$}{E-mail: xue@milano.infn.it}

\author {\bf Giuliano Preparata$^{a,b}$ and She-Sheng Xue$^{a,\dagger}$}
{\vbox{
\centerline{a) INFN, Section of Milan, Via Celoria 16, Milan Italy}
\centerline{b) Physics Department, University of Milan, Italy}}}

\abstract

We analyse the problem of the generation of mass in the gauge boson sector
within
our formulation of the Standard Model on a Planck lattice. We obtain a
dynamical scenario quite similar to the popular Higgs-mechanism, with the
important difference that no Higgs-type excitation is present in our theory.
By fixing the mass of the $Z^\circ$-boson and the renormalized gauge couplings
at their experimental values, we get for the top quark mass $m_t\simeq=149$GeV
and for the $\rho$-parameter the very small value $\rho\sim 10^{-6}$.

Proceeding on our research program which aims at formulating the Standard Model
(SM) on a Planck lattice (PL), i.e.~on a simple cubic lattice whose lattice
constant $a$ is just the Planck length, $a_p\simeq 10^{-33}$cm, in this letter
we wish to report on the preliminary investigation of the electroweak sector
$[SU(2)_L\otimes U(1)_Y]$, focusing on the propagation properties of the four
gauge
bosons: $W^\pm, Z^\circ$ and the photon.

The general formulation as well as the main motivations of the Standard Model
on the Planck lattice have been exposed in rather extensive detail in a recent
paper, to which we refer the interested reader\citref{xue92}. Here we only
remark that our research program stems from Wheeler's suggestion\citref{planck}
that at the Planck length $a_p$ space-time dissolves into the foam of the
violent
quantum fluctuations of the gravitation field, becoming a discrete structure
that can be usefully modelized as a Planck lattice. This view, that would
appear
as mainly metaphysical, when embodied in a real physical lattice gauge theory
has produced a number of surprising and satisfactory results, that
paradoxically
would seem to bring the extremely remote Planck length much closer to the
physical reality we have access to.

The most important aspect, to our opinion, that has so far emerged from
recent researches\citref{smit}\citref{rome} and our analysis
\citref{xue92}\citref{xue91}
is that the difficulty, the ``no-go'' theorem, that Nielsen and Ninomiya
\citref{nogo} had found at the beginning of the eighties in the straightforward
formulation of the SM on any lattice\footnote{(*)}{This means that even for
the usual continuum SM one does not possess a believable, gauge-invariant
means of introducing an ultraviolet cut-off.}, turns out to be a powerful
indication that something in the SM is {\it really missing}, and that such a
missing element is most likely a 4-fermion interaction of the Nambu-Jona
Lasinio (NJL) type\citref{nambu1}, that evades {\it in principle} the ``no-go''
theorem.

Our starting point is the PL (Euclidean) action
$$
S_{PL} = S_G
+ \sum_F \left(
S^F_D + S^{F_1}_{NJL} + S^{F_2}_{NJL}
\right),\eqnnum{spl}
$$
where $S_G$ $(S_D)$ is the usual SM gauge (Dirac) action on a lattice,
$$
S^{F_1} _{NJL} = -G_1 \sum_{x} \bar \psi^F_L (x)^{il} \cdot \psi^F_R (x)^{jk}
\bar \psi^F_R (x)^{jk} \cdot \psi^F_L (x) ^{il}
\eqnnum{s1}
$$
and
$$
\eqalign{
S^{F_2}_{NJL} &= -{G_2 \over 2} \sum_{x, \pm \mu} \big[ \bar \psi^F_L (x)
L^F _\mu (x) U^F_\mu (x) \cdot \psi^F_R (x + a_\mu) \cr
& \bar \psi^F_R (x) \cdot R^F _\mu(x) U^F _\mu(x) \psi ^F_L (x + a_\mu)\big],
}\eqnnum{s2}
$$
where the indices $F=q,l$ of the Fermi fields $\psi^F_{il}(x)$ denote the
quark and the lepton sector respectively; $i$ indexes the flavours and $l$ the
weak isospin. The dot ``$\cdot$'' denotes the Dirac color inner product, the
chiral gauge links
$L^F_\mu(x) (R^F_\mu(x))$ connect $\bar \psi ^F_L(x)$ and $
\psi^F_L(x+a_\mu) (\bar \psi^F_R(x)$ and $\psi^F_R(x+a_\mu))$, while the colour
links $U^c_\mu(x)$ pair up $\bar \psi^F_L(x)$ with $\psi^F_R(x +
a_\mu)$ and $\bar \psi^F_R(x)$ with $\psi^F_L(x+a_\mu)$. $G_1$
and $G_2$ are two new Fermi couplings $O(a^2)$.
$U^c _\mu(x)\in SU_c(3)$ appears in the quark-sector only. The chiral gauge
links $L^F _\mu(x)$ and $ R^F _\mu(x)$ are
$$
[R^F _\mu(x)]_{ij} =\delta_{ij}\left(\matrix{V_\mu(x)^{Y^F_{R_1}} & 0\cr
 0 & V_\mu(x)^{Y^F_{R_2}}}\right)\hskip0.5cm
\left[ L^F _\mu(x)\right]^{ij}_{kl} = \delta^{ij} \left[ U^L_\mu(x)
V_\mu(x)^{Y^F_{L}}\right]_{kl}
,\eqnnum{link}
$$
where $U^L _\mu(x)\in SU_L(2)$ and $V_\mu(x)\in U_Y(1)$,
and the hypercharge $Y^F$ has the usual assignments.

{}From the way it has been constructed the action\citeqn{spl} possesses in
addition to the gauge symmetries of the SM,
$SU_c(3)\otimes SU_L(2)\otimes U_Y(1)$, a new $U(N_g)\otimes U(N_g)$ global
symmetry where $N_g$ is the number of fermion families. As discussed
in\citref{xue92}, turning off the gauge degrees of freedom in the
action\citeqn{spl}, one finds that the two NJL-terms we have added to the
usual SM action lead to non-trivial ground states where:
\item{(i)} the troublesome doublers can be eliminated by the Wilson
mechanism\citref{wilson} through the emergence of a non-zero (O(1)):
$$
(\bar r^F)^{jk}_{il} = {G_2 \over 8V_4}
\sum_{\pm \mu,x} \left[\vev{ \bar \psi^F_L(x)^{jk} U^F_\mu(x) \psi^F_R (x +
a_\mu)_{il}} +\hc\right]
,\eqnnum{wilson}
$$
for every fermion;
\item{(ii)} the mass matrices:
$$
(M^F)^{jk}_{il} = - {G_1 \over 2V_4}
\sum_x \left[\vev{\bar \psi_L^F(x)^{jk} \psi_R^F(x)_{il}}+\hc\right],
\eqnnum{mass}
$$
contain ``mass counterterms''\citref{rome}, and physical masses $m_F$ which
have non-zero elements for one family of quarks only, to be identified with
$(t, b)$. All other quarks and leptons remain at this stage massless;
\item{(iii)} there appear in the spectrum four massless Goldstone bosons, whose
composite fields are
$$
\eqalign{
\phi^-(x)&=\vev{\bar t(x)\gamma_5b(x)},
\hskip1cm \phi^+(x)=(\phi^-(x))^\dagger,\cr
\phi^\circ(x)&={1\over\sqrt{2}}\big[\vev{\bar t(x)\gamma_5 t(x)}-
\vev{\bar b(x)\gamma_5 b(x)}\big],\cr
\bar\phi^\circ(x)&={1\over\sqrt{2}}\big[\vev{\bar t(x)\gamma_5t(x)}+
\vev{\bar b(x)\gamma_5b(x)}\big];}\eqnnum{gold}
$$
\item{(iv)} the only gauge symmetry left is $SU(3)_c\otimes U(1)_{em}$, and the
scalar composite (the Higgs candidates\citref{Nambu}\citref{bardeen})acquires
a mass of the order $m_p\simeq 10^{19}\gev$, thus disappearing from the low
energy spectrum.

We now turn on the gauge interaction and consider the complete
action $S_{PL}$. The partition function $Z$ can be written (somewhat
impressionistically)
$$
Z=\int [d\psi d\bar\psi][dLdR]
e^{-S_{PL}(\psi,\bar\psi,L,R)};\eqnnum{parti}
$$
the non-trivial vacuum, which as we have discovered at the first stage ( where
the gauge fields were turned off) is
characterized by non vanishing $r$ and $M$ (see eqs.\citeqn{wilson}
and\citeqn{mass}), allows us to build the effective (broken) theory upon the
partition function,
$$
\eqalign{
Z(M,r)&=\int [dLdR][d\psi d\bar\psi]
e^{-S_{PL}(\psi,\bar\psi,L,R)}\cr
&\left[\delta\left(-{G_1\over 2V_4}\sum_x\bar \psi_L^F(x)
\psi_R^F(x)-M^F\right)
\cdot\delta\left({G_2\over 8V_4}\sum_{\pm\mu,x}\bar \psi^F_L(x) U^F_\mu(x)
\psi^F_R (x + a_\mu)-r^F\right)\right].}\eqnnum{eff}
$$
A relevant observation at this point is that, due to the gauge invariance of
$S_{PL}$, $Z(M,r)$ is invariant under the class of local gauge transformations
that leave invariant the arguments of the $\delta$-functions appearing
in\citeqn{eff}. It is easy to check that such gauge transformations,
belonging to the groups $SU_L(2)\otimes U_Y(1)$, are of the type
$U_L(x)=\exp[i\alpha^i(x){\tau^i\over 2}], U_Y(x)=\exp[i\alpha_0(x)]$, where
the four
fields $\alpha_i(x)$ and $\alpha_0(x)$ do not possess zero-modes in
momentum space, i.e.~ for low momenta they behave like massless fields. It is
then natural to identify the appropriate linear combinations of such fields
with the Goldstone fields\citeqn{gold} that from our previous
analysis\citref{xue92} we have found to be associated with the phenomenon of
mass generation.
By setting $\psi^F(x)=\psi^F_0(x)+\psi^F_s(x)$, where
$\psi^F_0(x)=U(x)\Psi^F_0, U=U_L, U_Y$ and $\Psi^F_0$ are constant
Dirac spinors obeying $\bar\Psi^F_0\Psi^F_0\sim M_F, r^F$, we can
rewrite the partition function\citeqn{eff} as
$$
Z(M,r)=\int [d\psi_s d\bar\psi_s][dLdR][dU_LdU_Y]
e^{-S_{PL}(\psi_s,\bar\psi_s,M,r,L',R')},\eqnnum{effpart}
$$
where $L_\mu'$ and $R_\mu'$ are the gauge links resulting from the
transformations
$U_L(x)$ and $U_Y(x)$.
It can be easily ascertained that the
primed gauge links $L_\mu'(x)$ and $R'_\mu(x)$ add to the gauge-fields
$W^\pm_\mu, Z^\circ_\mu$ and $A^{em}_\mu$ ( we use the standard nomenclature)
the terms $-{1\over g_2}\partial_\mu \phi^\pm(x),
-{1\over\beta}\partial_\mu\phi^\circ(x)$ and
$-{1\over Q}\partial_\mu\bar\phi^\circ$
respectively ( $g_2$ is the $W^\pm$ coupling, $\beta$ is the neutral current
coupling of the $Z^\circ$ and $Q$ is the electric charge). This is just
what we wanted, actually needed, for the three Goldstone modes $\phi^\pm$
and $\phi^0$ become, as in the familiar Higgs mechanism, the longitudinal modes
of the weak gauge bosons $W^\pm$ and $Z^\circ$. As for the unwanted fourth
Goldstone boson $\bar\phi^\circ$, the remaining unbroken symmetry $U(1)_{em}$
guarantees that it can be gauged away, becoming an invisible phase of the
fermion fields.

By integrating over the fermion fields we get,
$$
Z(M,r)=\int [dLdR][dU_LdU_Y]
e^{-V(M,r,L',R')},\eqnnum{effpot}
$$
where the effective potential can be expanded as:
$$
V(M,r,L',R')=\sum^\infty_0\Gamma^{(n)}[L',R',M,r],\eqnnum{gam}
$$
$\Gamma^{(n)}$ being the n-point one-particle irreducible Green functions.
In the final part of this paper we shall focus our attention only on
$\Gamma^{(2)}(L',R',M,r)$, i.e.~on the inverse propagators of the electroweak
gauge bosons $W_\mu^\pm, Z^\circ_\mu$ and $A^{em}_\mu$.

We have noticed above that, as far as gauge invariance is concerned, the
only difference between the gauge invariant partition function\citeqn{parti}
and $Z(M,r)$ is that the latter is invariant for all gauge transformations
that leave conditions\citeqn{wilson} and\citeqn{mass} invariant, i.e.~are
characterized
by transformation functions that do not have zero modes in momentum space.
This means that, if we can limit ourselves to these latter gauge transformation
functions, $Z(M,r)$ has the same gauge-invariance of the original symmetric
theory. This general argument is enough to ensure that $\Gamma^{(2)}(L',R')=
\Gamma^{(2)}(L,R)$, i.e.~the inverse propagators for all gauge-fields
are purely transverse\citref{xuewill}.

Getting now to the actual gauge-boson self-energy function, that we may
write as
$$
\Pi_{\mu\nu}(k)=\big({k_\mu k_\nu\over k^2}-\delta_{\mu\nu}\big)\Pi(k^2),
\eqnnum{prop}
$$
this is given by the diagrams in \Fig{ds},
where the fermion propagator $-----$ satisfies the mean field gap equation
of Ref.\citref{xue92}; the $(b)$-terms come from the Dirac action $S_D$,
while the $(c)$-contributions stem from the composite modes (both scalar and
Goldstone). The terms in the $(d)$ class are typical lattice contributions
\citref{rome} originating from the two NJL-terms of the action. In order to
isolate the transverse part $\Pi(k^2)$ (all non-transverse terms
by virtue of the preceding general argument must vanish
identically) we divide the integration region $[-\pi,\pi]^4$ of the internal
loop momenta $l_\mu=ap_\mu$ in the ``continuum region'' $|l_\mu|\le\epsilon$
and the ``lattice region'' $|l_\mu|\ge\epsilon$, with $a_p|k_\mu|\ll\epsilon
\ll\pi$. The contribution to $\Pi(k)$ from the lattice region of the diagrams
of types (b) (c) can be written:
$$
\eqalign{
\Pi^l(k)&=\sum_F2N_F(\beta^2_L+\beta^2_R)
\big[c_1(r_F)-{1\over 48\pi^2}\ln\epsilon\big]\cr
&+2\sum_F(\beta_L-\beta_R)^2_Fm_F^2N_F\big[c_2(r_F)-{1\over 8\pi^2}
\ln\epsilon\big]
,}
\eqnnum{lattice}
$$
where $\beta_{L,R}$ are the left- and right-handed couplings of the given
gauge-bosons
respectively, $N_F=3(0)$ for quarks(leptons) and $c_{1,2}(r_F)$ are very small
functions of $r_F$ that are determined numerically.

In the continuum region the contributions $(a),(b)$ and $(c)$ are exactly the
same
one computes in continuum field theory. One gets,
$$
\Pi^c(k)=k^2\big[1+\sum_F\pi^F_\Lambda (k)\big]+\sum_Fm^2_F\mu^F_\Lambda (k),
\eqnnum{tran}
$$
where
$$
\pi^F_\Lambda (k)=2(\beta^2_L+\beta^2_R)_FN_F\int_0^1dx(1-x)x\int_\Lambda
{d^4p\over
(2\pi)^4}{1\over [p^2+m_F^2+k^2x(1-x)]^2}\eqnnum{nomass}
$$
and
$$
\mu^F_\Lambda(k)=2(\beta_L-\beta_R)^2_FN_F \int_0^1dx\int_\Lambda {d^4p\over
(2\pi)^4}{1\over [p^2+m_F^2+k^2x(1-x)]^2}.\eqnnum{massfun}
$$
Note that $\Lambda=\epsilon\Lambda_p$,
where $\Lambda_p={\pi\over a_p}\simeq10^{19}$GeV.
Combining $\Pi^l(k^2)$ with $\Pi^c(k^2)$ we first notice that the arbitrary
$\epsilon-$
scale, dividing the continuum region from the lattice region, disappears as it
should, giving for the masses of the gauge bosons ($k^2=-M_G^2$):
$$
M_G^2=2\sum_Fm_F^2(\beta_L-\beta_R)^{2,rem}_FN_F
\Big[\int_0^1dx\int_{\Lambda_p}
 {d^4p\over
(2\pi)^4}{1\over [p^2+m_F^2-M_G^2x(1-x)]^2}+c_2(r_F)\Big],\eqnnum{gmass}
$$
where the renormalized gauge couplings are
$$
\beta^{2,rem}_{L,R}=Z_3(k^2=-M_G^2)\beta^2_{L,R};\hskip1cm Z^{-1}_3(k^2)=1+
\sum_F\big[\pi^F_{\Lambda_p}(k)+\bar c_1(r_F)\big],\eqnnum{reno}
$$
and $\bar c_1(r_F)=2N_F(\beta_L^2+\beta_R^2)c_1(r_F)$. Eqs.\citeqn{gmass}
and\citeqn{reno} are the final results of our calculation. The
mass of the photon, due to $\beta^F_L=\beta^F_R=Q_F$, pleasingly turns out to
vanish, as it obviously should. As for the $Z^\circ$, taking $m_t\not=m_b$ and
neglecting the masses of the other fermions, we
have:
$$
M_z^2={3\alpha\over 8\pi\sin^2\theta_w\cos^2\theta_w}\big[m_t^2f(m_t)_z+
m_b^2f(m_b)_z],\eqnnum{zmass}
$$
where
$$
f(m_F)_G=\int_0^1dx\int_{\Lambda_p} {d^4p\over
(2\pi)^4}{1\over [p^2+m_F^2-M_G^2x(1-x)]^2}+c_2(r_F).\eqnnum{simass}
$$
Setting $m_b\simeq 5$GeV, $M_z\simeq 91.2$GeV, $\alpha={1\over137},
\sin^2\theta_w=0.23$ and neglecting $c_2(r_F)\ll 1$ we obtain the prediction
\footnote{(*)}{This prediction of $m_t$ agrees with the one we have recently
made in Ref.\citref{xue93} based on the Dyson equations of our theory relating
$m_t$ to the bottom quark mass $m_b\simeq 5GeV$, as determined from quark
spectroscopy.}
$$
m_t=1.633M_z=149\gev.\eqnnum{top}
$$
Concerning the W-mass $M_w$, setting as usual
$$
{M^2_w\over M^2_z}=\cos^2\theta_w(1+\rho),\eqnnum{ratio}
$$
$\rho$ is given by
$$
\rho=\left({m_b\over m_t}\right)^2\left[{f(m_b)_w\over f(m_t)_w}-
{f(m_b)_z\over f(m_t)_z}\right]\simeq \left({m_b\over m_t}\right)^2{\ln\left({
M_z^2\over M^2_w}\right)\over \ln\left({
\Lambda_p^2\over m^2_t}\right)}\simeq 10^{-6},\eqnnum{rho}
$$
a very small value, consistent with experiments.

Let us conclude this letter with a brief discussion/assessment of what
we have been able to achieve. As already stressed, this is the first paper
in our
research program in which the dynamics of the gauge bosons has been analysed,
with the aim of getting a detailed picture of the fundamental mechanism by
which gauge bosons get their masses. Our previous
work, in particular the discovery of the four Goldstone modes associated with
the generation of the masses of the top and bottom quarks, did make us expect
for the masses of the gauge bosons a scenario of the type investigated by Nambu
\citref{Nambu} and Bardeen and collaborators\citref{bardeen}, that had already
exhibited many satisfactory features. Our expectations have been totally
confirmed, as one can check by looking at eqs.\citeqn{top} and\citeqn{rho}.
But there is more, our general analysis of the gauge invariance of the
partition
function $Z(M,r)$ demonstrates that the ultraviolet properties of our theory
are completely under control, leading to a coupling between the continuum
region
(where we presently live) and the Planck lattice region that is only
{\it logarithmic}, thus leaving only a feeble echo of the violent quantum
fluctuations, which in our view tear space-time apart, making it discrete
structure.

Barring unpleasant surprises, we expect that our Planck lattice SM will
produce a theory of the electroweak phenomena that, without low mass
($m_H\ll \Lambda_p$) Higgs fields, reproduces all the dynamics that has so far
been
associated to the Higgs-mechanism. At higher energies, i.e.~in the TeV region,
the theory should preserve its perturbative, unitary structure by cutting off
the unwanted longitudinal amplitudes at a mass scale comparable with $m_t$ or
$M_{w,z}$, and not with the mass the non-existent (in our approach) Higgs
meson.
But this will be the object of the next stage of our research program.

\newref{nambu1}
Y.~Nambu and G.~Jona-Lasinio, \journal \PR 122,61,345.

\newref{Nambu}
Y.~Nambu, in New theories in physics, Proc.\ XI Int.\ Symp.\ on elementary
particle physics, eds.~Z.~Ajduk, S.~Pokorski and A.~Trautman (World Scientific,
Singapore,~1989).

\newref{bardeen}
W.A.~Bardeen, C.T.~Hill and M.~Linder, \journal \PR D41,90,1647;\addref
V.A.~Miranski, M.~Tanabashi and K.~Yamawaki, \journal \MPL A4,89,1043;
\journal \PL B221,89,117.

\newref{xue91}
G.~Preparata and S.-S.~Xue, \journal \PL B264,91,35; \addjour B302,93,442;
MITH 92/1 submitted to
\MPL; \journal \NP B26,92,501 (Proc.~Suppl.);
\journal \NP B30,93,674 (Proc.~Suppl.).

\newref{nogo}
H.B.~Nielsen and M.~Ninomiya, \journal \NP B185,81,20; \addjour B193,81,173;
\journal \PL B105,81,219.

\newref{planck}
C.W.~Misner, K.S.~Thorne and J.A.~Wheeler, {\it Gravitation\/} (Freeman,
San Fransisco, 1973).

\newref{smit}
J.~Smit, \journal {Acta Physica Polonica} B17,86,531;\addref
P.D.V.~Swift, \journal \PL B145,84,256;\addref
S.~Aoki, \journal \PRL 60,88,2109; \journal \PR D38,88,618;\addref
I.~Montvay, \journal \PL B199,87,89; \journal \PR B205,88,315;\addref
E.~Eichten and J.~Preskill, \journal \NP B268,86,179;\addref
K.~Funakubo and T.~Kashiwa, \journal \PRL 60,88,2113;
\journal \PR D38,88,2602;\addref
S.~Aoki, I.~H.~Lee and S.-S.~Xue, \journal \PL B229,89,77\addref
T.~D.~Kieu, D.~Sen and S.-S.~Xue, \journal \PRL 61,88,282\addref
S.-S.~Xue, \journal \PL B245,90,565.

\newref{rome}
A.~Borrelli, L.~Maiani, G.C.~Rossi, R.~Sisto and M.~Testa,
\journal \NP B333,90,335; \journal \PL B221,89,360.

\newref{xuewill}
We shall discuss this problem in a separate paper.

\newref{xue93}
G.~Preparata and S.-S.~Xue, ``Mass relation between top and bottom quarks'',
MITH 93/6, {\it to appear Phys.~Lett.~B}.

\newref{xue92}
G.~Preparata and S.-S.~Xue, `` Emergence of the $\bar tt-$condensate and
the disappearance of Scalars'', MITH 93/5.

\newref {wilson}
K.~Wilson, in {\it New phenomena in subnuclear physics\/}
(Erice, 1975)
ed.\ A.~Zichichi (Plenum, New York, 1977).

\newfig {ds} {The diagrammatic form of Dyson-Schwinger equation for
$\Pi_{\mu\nu}(k)$. }
